\documentclass[pre,twocolumn,showpacs,showkeys,preprintnumbers,amsmath,amssymb]{revtex4-1}
\usepackage{amssymb}
\usepackage{graphicx}
\usepackage{array}
\usepackage{hhline}
\usepackage{longtable}
\begin{document}
\title{Dynamically order-disorder transition in triangular lattice driven by a time dependent magnetic field}
\author{E. Vatansever}
\email{erol.vatansever@deu.edu.tr}
\affiliation{Department of Physics, Dokuz Eyl\"{u}l University, TR-35160, Izmir-Turkey}

\date{\today}
\begin{abstract}
We have elucidated the dynamic phase
transition features and finite-size scaling analysis of the triangular lattice
system under the presence of a square-wave magnetic field. It has been found that
as the value of half-period of the external field reaches its critical value, whose location
is estimated by means of Binder cumulant, the system presents a dynamic phase transition between
dynamically  ordered and disordered phases.  Moreover, at the dynamic phase
transition point, finite-size scaling of the Monte Carlo
results  for the dynamic order  parameter and susceptibility give the critical
exponents $\beta/\nu=0.143\pm0.004$ and $\gamma/\nu=1.766\pm0.036$, respectively. The obtained critical exponents show that present magnetic
system belongs to same universality class with the two-dimensional equilibrium Ising model.
\end{abstract}
\maketitle
\section{Introduction}\label{Introduction}
The physical mechanism behind nonequilibrium phase transitions is less
understood than that of equilibrium phase transitions for magnetic systems,
and it deserves particular attention. Interacting spin systems
under the existence of  an oscillating magnetic field can display unusual and
interesting magnetic behaviors, which can not be observed in their
corresponding equilibrium parts. For the first time, the authors in Ref. \cite{Tome}
applied their mean field tools to characterize the kinetic nature of the Ising model
being subjected to a time dependent magnetic field. From their analysis, it has been
found that amplitude and period of the external field have an important role
on the dynamic behavior of the studied system. For example, the system undergoes a dynamic phase
transition (DPT) between dynamically ordered and disordered phase with increasing value
of the applied field amplitude by keeping other system parameters fixed. Since then, many
theoretical \cite{Lo, Zimmer, Acharyya1, Acharyya2, Acharyya3, Acharyya4, Jang1, Jang2, Jang3, Chakrabarti,
Shi, Keskin, Punya, Yuksel1,  Vatansever} and several experimental \cite{He, Robb, Suen, Berger, Riego1}  studies
have been performed to examine the  DPTs and to understand in  depth
their origins observed in different  types of magnetic systems. Note that in most of the
theoretical studies mentioned above, DPT have been encountered by changing the
applied field amplitude and temperature.

Some efforts were also taken to elucidate the influences of the period of the external
magnetic field on the dynamic phase transition phenomena at constant
applied field amplitude \cite{Sides1, Sides2, Korniss1, Buendia, Park1, Park2,
Idigoras, Tauscher, Riego2}.  Below its equilibrium critical temperature $T_{c}$,
the kinetic Ising model undergoes a DPT between dynamically ordered and
disordered phase when the period of the field reaches the critical period. For small period
values of the field, the system does not have enough time to follow the external field
instantaneously. Thereby, time dependent magnetization oscillates around a non-zero
value indicating a dynamically ordered phase. However, magnetization can be capable
of following the external field with a relatively small phase lag, which indicates the
dynamically disordered phase. Some of previously published
works indicate that  there is a good consensus between DPTs
and equilibrium phase transitions, especially for the determination of universality
class of the spin system far from equilibrium. For instance, it has been found that
the critical exponents of the two-dimensional (2D) kinetic Ising model subjected to a
square-wave oscillatory magnetic field are consistent with the universality class of the
corresponding 2D equilibrium Ising model \cite{Sides1, Sides2, Korniss1,
Buendia}. In another report,  finite-size scaling analysis of Monte Carlo simulation
supports these findings for the three dimensional kinetic Ising model \cite{Park1}.
These studies also  show that the   symmetry arguments reported in Ref. \cite{Grinstein} is valid
for the magnetic systems without surfaces \cite{Park2} driven by a time dependent
external field.  We would like to mention that particular interests in works discussed
above  have been only dedicated to classify the universality classes of the
square and simple cubic lattices in detail. We believe that
much more work is required to have better understanding of the DPTs
and classifying universality properties of the spin systems far from
equilibrium in different geometries such as triangular, honeycomb,
and kagome lattices.

In the present work, we consider the kinetic Ising model on a triangular lattice being
subjected to a square-wave magnetic field, in order to contribute to the finite-size
scaling properties and also universality properties of spin system far from
equilibrium. Based on the finite-size scaling of the Monte Carlo results for the
dynamic order parameter and susceptibility, it has been estimated the
critical exponents.  The obtained critical exponents demonstrate that present magnetic
system belongs to same universality class with the 2D
equilibrium Ising model.

The outline of remainder parts of the paper is as follows: In section \ref{Model},
we give details of model and simulation procedure. The results and discussion
are presented in section \ref{Discussion}, and finally
section \ref{Conclusion} contains our conclusions.

\section{Model and Simulation Details}\label{Model}
We study the kinetic Ising model on a triangular lattice under
presence of a time dependent magnetic field. The Hamiltonian of the present system
can be written as follows:
\begin{equation}\label{Eq1}
H=-J\sum_{\langle ij \rangle} S_{i}S_{j}-h(t)\sum_{i}S_{i}
\end{equation}
where $S_{i}=\pm 1$ is the Ising spin variable at the position $i$, and $J$ is the
ferromagnetic $(J>0)$ spin-spin coupling between nearest neighbor $(nn)$ spins in
the system. The first summation  in Eq. (\ref{Eq1}) is over the $nn$ site pairs in
the system while the second one is over the all lattice  sites in the 2D triangular
lattice system. $h(t)$ describes the time dependent oscillating magnetic field.
For the present study, we use a square-wave magnetic field source with amplitude $h_{0}$
and half-period $t_{1/2}$, following the references \cite{Korniss1, Buendia, Park1}.

We use Monte Carlo simulation with local update Metropolis algorithm \cite{Binder, Newman}
to understand and clarify the DPT characteristics
and universality properties of the  system on a $L\times L$ triangular
lattice, where $L$ is the linear size of the system.
Periodic  boundary conditions are applied to the system in all directions.
We consider the initial configuration where all spins are up, and
spin configurations are generated by selecting the lattice
site randomly through the triangular lattice.
Here, we restrict  ourselves to consider the values of the
field amplitude $h_{0}/J=0.3$ and of the  temperature $T=0.8T_{c}$,
where $T_c=3.60495J/k_{B}$ is the critical temperature of the 2D
triangular lattice Ising model. After discarding the first 1000
period of the external field, numerical data were collected
over next 200 000 periods of the field. We note that the time unit is one
Monte Carlo step per site (MCSS).

In order to elucidate the critical properties of the
dynamic phase transitions, one can consider the dynamic order
parameter, which is the time  averaged magnetization over a full
cycle of the external magnetic field:

\begin{equation}\label{Eq2}
Q=\frac{1}{2t_{1/2}}\oint M(t)dt,
\end{equation}
here $M(t)$ is the instantaneous value of the magnetization
per site, which can be obtained as follows:
\begin{equation}\label{Eq3}
M(t)=\frac{1}{L^2}\sum_{i=1}^{L^2}S_{i}.
\end{equation}
We note that due to the symmetry of the system, the probability distribution of the dynamic order parameter
is bimodal form in the dynamically ordered phase for the finite lattice sizes.
Keeping this in mind, the order parameter is considered
as $\langle |Q| \rangle$, namely average norm of $Q$.

In order to determine the dynamic critical point with a high precision,
one of the  suitable ways is to calculate  Binder cumulant as a
function of the system size:

\begin{equation}\label{Eq4}
U_{L}=1-\frac{\langle Q^4 \rangle_L}{3\langle Q^2\rangle_L ^2}.
\end{equation}

Previous studies on the universality aspects of the kinetic Ising model suggest
that the scaled variance of the dynamic order parameter can be regarded as susceptibility of
the system, which can be defined as follows:

\begin{equation}\label{Eq5}
\chi^{Q}_{L}=L^2\left(\langle Q^2\rangle-\langle |Q|\rangle^2 \right).
\end{equation}
In order to extract the critical exponents, one of the well-known
methods is finite-size scaling method. In this method, the main tool is to
determine the measured quantities as a function of the system size. Based on
the finite-size scaling method for the system in thermal
equilibrium \cite{Privman1, Binder, Newman}, it is possible
to write down the following scaling forms for the order parameter
and susceptibility at the critical point:
\begin{equation}\label{Eq6}
 \langle |Q| \rangle_{L} \propto L^{-\beta/\nu},
\end{equation}

\begin{equation}\label{Eq7}
 \chi_{L}^Q \propto L^{\gamma/\nu}.
\end{equation}
Previous detailed investigations show that these scaling forms are also
applicable to classify the universality classes of the magnetic systems driven by a time
dependent oscillating magnetic field \cite{Sides1, Sides2, Korniss1, Buendia, Park1, Park2}.

\section{Results and Discussion}\label{Discussion}
\begin{figure}[h!]
\center
\includegraphics[width=6cm]{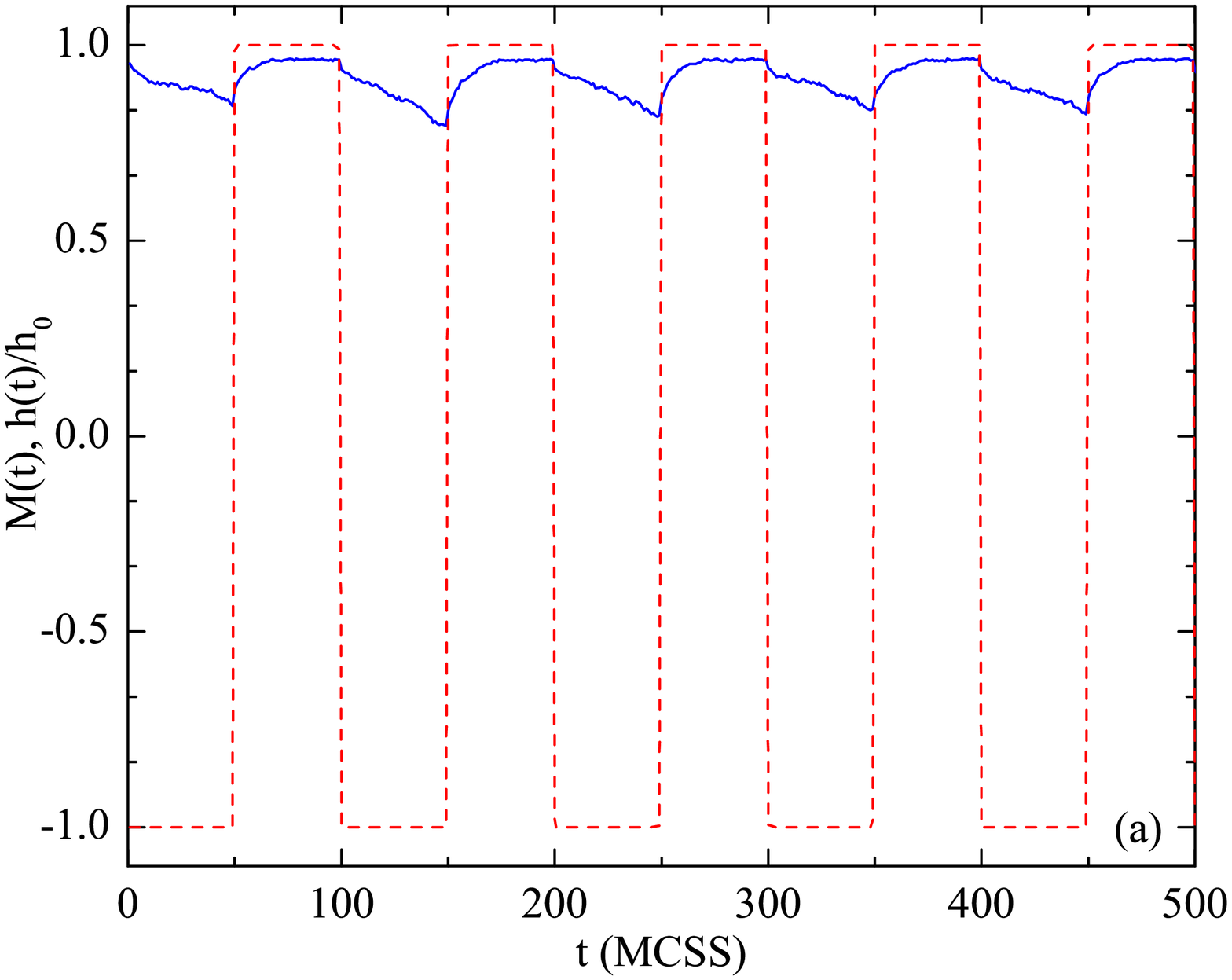}
\includegraphics[width=6cm]{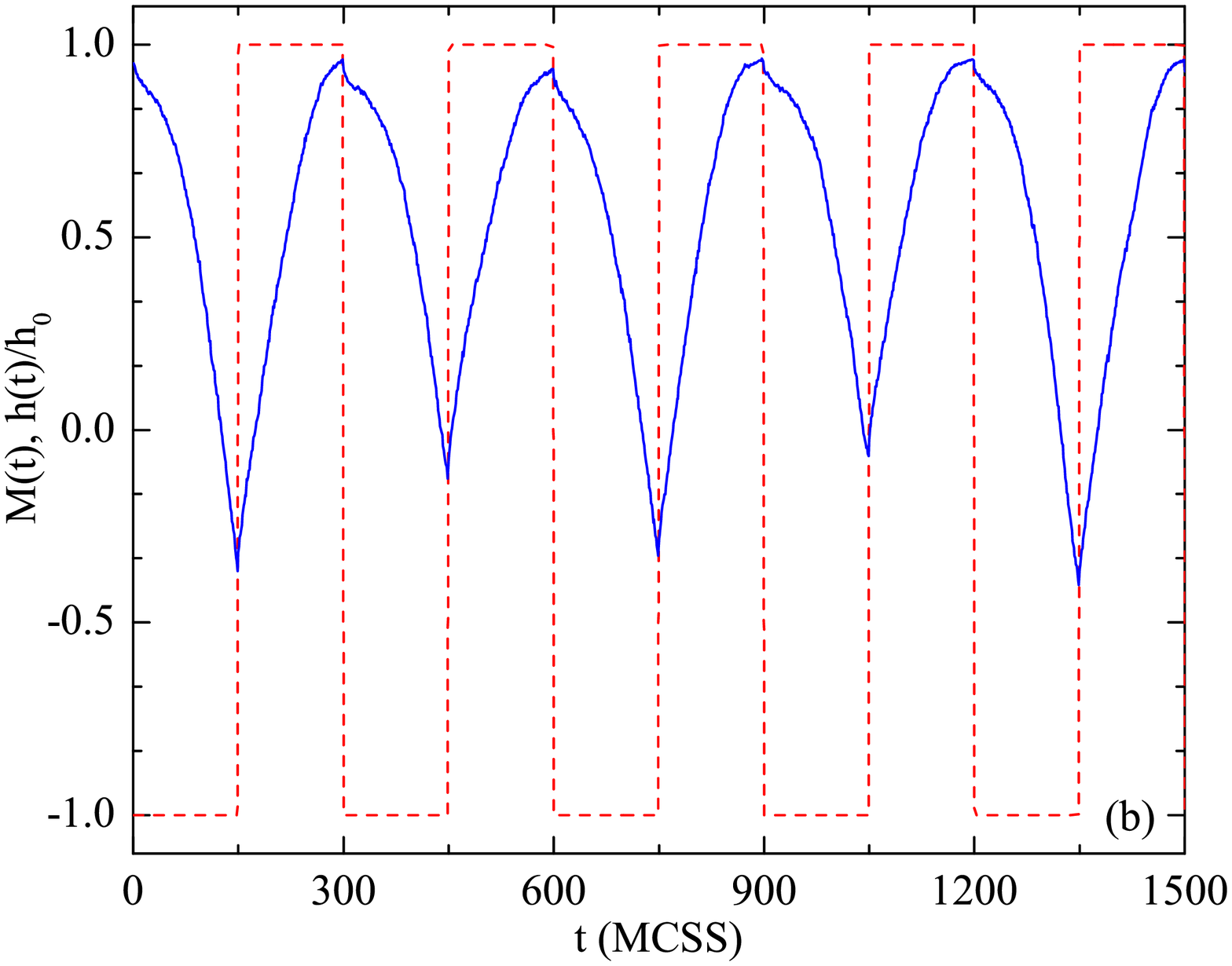}
\includegraphics[width=6cm]{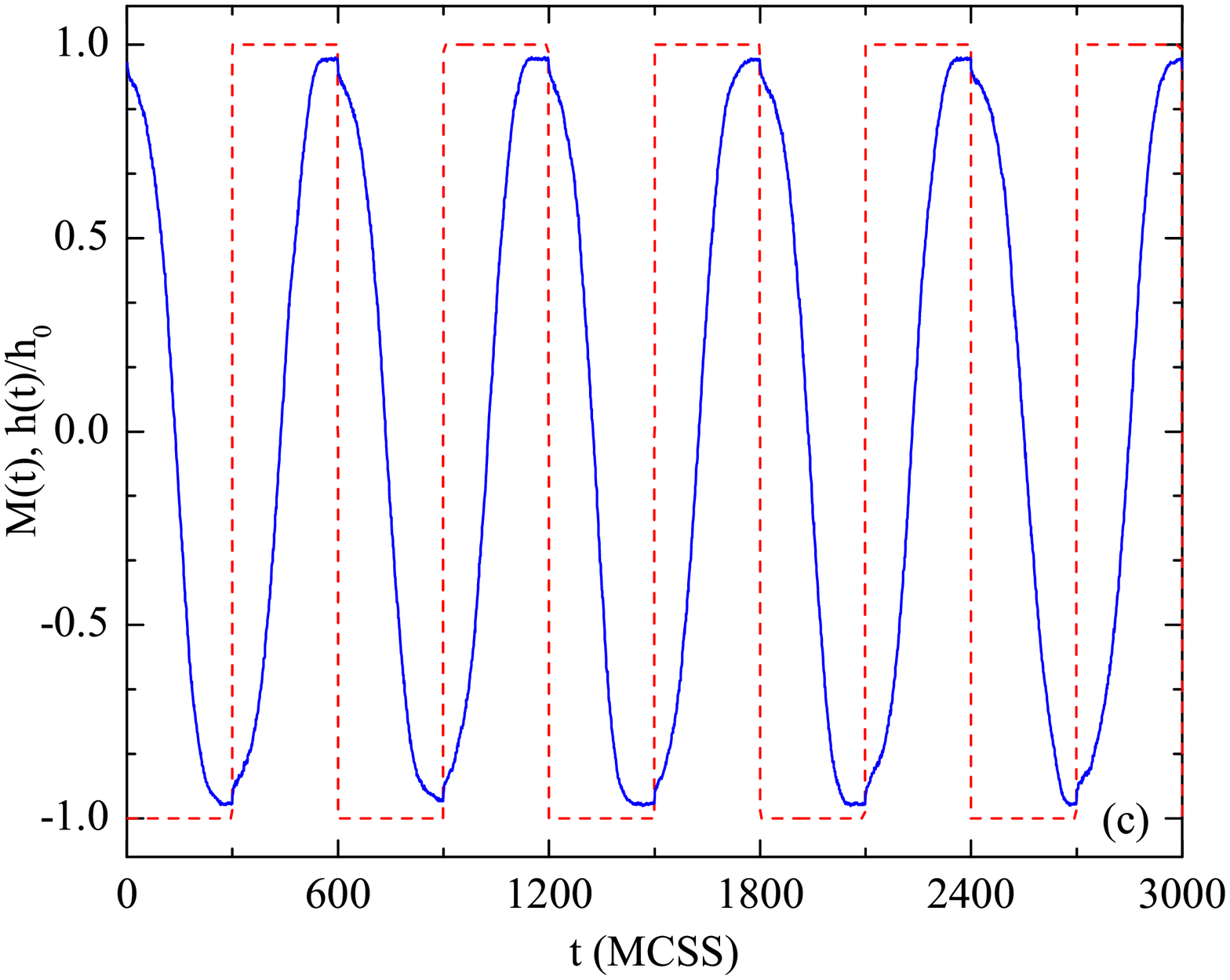}
\caption{(Color online) Time dependent magnetization (blue solid lines) of the
kinetic Ising model on a triangular lattice driven by a
square-wave magnetic field (red dashed lines denote $h(t)/h_0$
where $h_{0}$ is amplitude of field) for three considered values of the
half-period $t_{1/2}$ of the field. (a) $t_{1/2}=50$ MCSS and (c) $t_{1/2}=300$ MCSS correspond to the
dynamically ordered and disordered phases, respectively. (b) $t_{1/2}=150$ MCSS, it is close
to the dynamic phase transition point of the system. The numerical data were collected for a
system size $L=180$ at $T=0.8T_{c}$ and for value of $h_{0}/J=0.3$.}\label{Fig1}
\end{figure}
In Fig. \ref{Fig1}(a-c), we focus our attention on the time series of magnetization
of the kinetic Ising model on a triangular lattice for a
system size $L=180$ at $T=0.8T_{c}$ and $h_{0}/J=0.3$. The time series are plotted at various values of
the half-period of the external field: (a) $t_{1/2}=50$ MCSS, (b) $150$ MCSS and (c) $300$ MCSS, respectively. As shown
from the Fig. \ref{Fig1}(a), the magnetization of the system does not have enough time to follow
the rapidly changing external field. Thereby, it oscillates around a non-zero value corresponding to the
dynamically ordered phase $(Q\neq0)$. As the half-period of the external field is increased further,
for example $t_{1/2}=300$ MCSS, the system begins to reverse its magnetization corresponding to the dynamically
disordered phase $(Q=0)$, as shown in Fig. \ref{Fig1}(c). It is clear that there exists a
critical half-period value where a  DPT takes place. Our Monte
Carlo simulation results  suggest that the critical half-period of the external
field is $t_{1/2}^c=142\pm1$ MCSS (which will be discussed in the following) for the considered
kinetic Ising model on a triangular
lattice model. In Fig. \ref{Fig1}(b), we give an example of the time series of the magnetization in
the vicinity of  the DPT of the system for value of
half-period $t_{1/2}=150$ MCSS of the external applied magnetic field.

\begin{figure}[h!]
\center
\includegraphics[width=6cm]{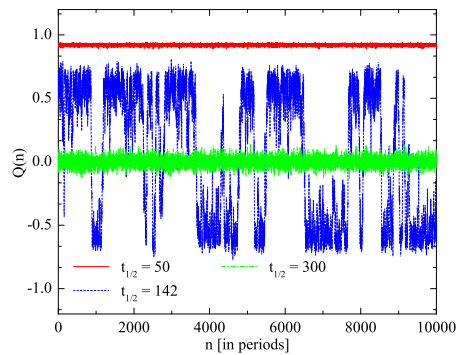}
\caption{(Color online) Period dependencies of the dynamic
order parameter $Q$ of the kinetic Ising model on a triangular
lattice, for the considered  system parameters $L=180$,
$T=0.8T_c$ and $h_{0}/J=0.3$. The curves are obtained for three values of
the half-period of the external field. $t_{1/2}=50$ MCSS corresponds to the
dynamically ordered phase where $Q$ oscillates around a finite
value. Dynamic order parameter exhibits strongly fluctuating behavior
at $t_{1/2}=142$ MCSS, indicating the existence of a DPT.
$Q$ oscillates around zero value for the value of $t_{1/2}=300$ MCSS of the
field, which is a signature of the dynamically disordered phase.}\label{Fig2}
\end{figure}

In Fig. \ref{Fig2}, we give period dependencies of the dynamic order parameter,
for the same system parameters used for  Fig. \ref{Fig1}. These curves are
demonstrated for three values of  the half-period of the external
field, i.e., $t_{1/2}=50, 142$ and $300$ MCSS.
For $t_{1/2}=50$ MCSS, the magnetic system exists in the dynamically ordered phase,
and hence $Q$ oscillates around a non-zero value. However, for $t_{1/2}=300$ MCSS, period
averaged $Q$ equals to zero indicating dynamically disordered phase. On the other hand, dynamic
order parameter displays strongly fluctuating behavior at $t_{1/2}^c=142$ MCSS.
Large fluctuation behavior observed in the $Q$ as a function of the period of the
external magnetic field is a clear evident of a DPT.

\begin{figure}[h!]
\center
\includegraphics[width=6cm]{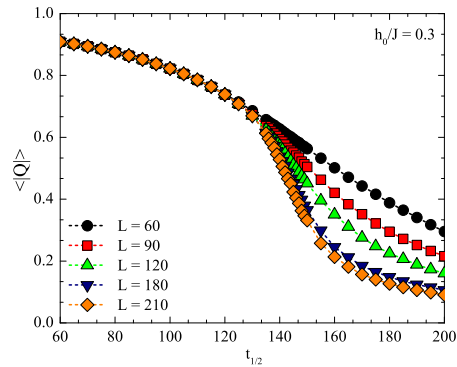}
\caption{(Color online) Half period dependency $t_{1/2}$ of
the dynamic order parameter $\langle |Q| \rangle_{L} $ of the
kinetic Ising model on a triangular lattice. The curves are obtained for varying
values of lattice sizes ranging
from $L=60$ to 210. The numerical data are collected by
averaging over 200 000 periods of the magnetic field.}\label{Fig3}
\end{figure}

\begin{figure}[h!]
\center
\includegraphics[width=6cm]{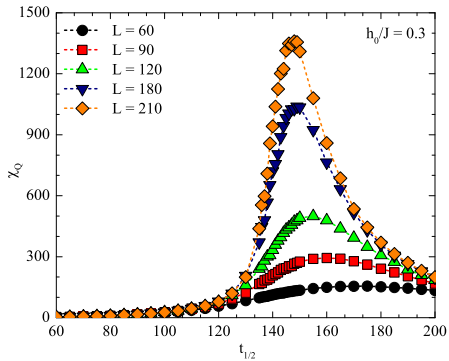}
\caption{(Color online) Half period dependency $t_{1/2}$ of
the dynamic susceptibility $\chi^{Q}$ of the
kinetic Ising model on a triangular lattice.
The $\chi^{Q}$ curves are obtained at various values of lattice sizes ranging
from $L=60$ to 210. The numerical data are collected by
averaging over 200 000 periods of the magnetic field.}\label{Fig4}
\end{figure}

In Figs. \ref{Fig3} and \ref{Fig4}, as an example of finite-size behavior, we show the data of the
dynamic order parameters and their fluctuations for various values of the lattice sizes ranging
from $L=60$ to 210. It is obvious from these figures that as value of the half-period of the
external field is increased starting from relatively lower values, dynamic order parameter begins to
decrease for all studied values of lattice sizes. We also note that half-period
dependency of $\langle|Q|\rangle$ tends to disappear with increasing system size. As displayed in Fig. \ref{Fig4},
their corresponding susceptibility curves represent a behavior which tends to diverge as the
lattice size of the system is increased,  in the neighborhood of DPT.

\begin{figure}[h!]
\center
\includegraphics[width=6cm]{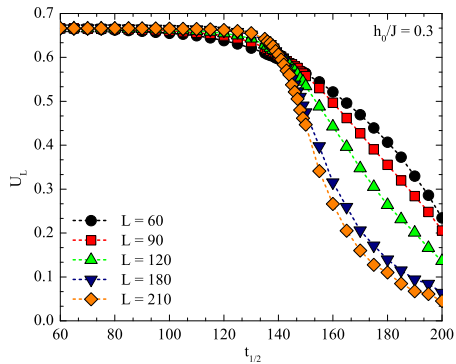}
\caption{(Color online) Half period dependency $t_{1/2}$ of
the Binder cumulant $U_{L}$ of the kinetic Ising model on a triangular lattice.
The curves are obtained at various values of lattice sizes ranging
from $L=60$ to $210$. The numerical data are collected by
averaging over 200 000 periods of the magnetic field. }\label{Fig5}
\end{figure}

In order to determine the critical half-period of the external field, we perform
half-period dependency $t_{1/2}$ of the Binder cumulant $U_{L}$ at varying values of
system size, as seen in Fig. \ref{Fig5}. Our Monte Carlo simulation results indicate
that the obtained Binder cumulants for varying lattice sizes cross at a special value of
half-period of the external field $t_{1/2}^c=142\pm1$ MCSS, where
DPT takes place.

\begin{figure}[h!]
\center
\includegraphics[width=6cm]{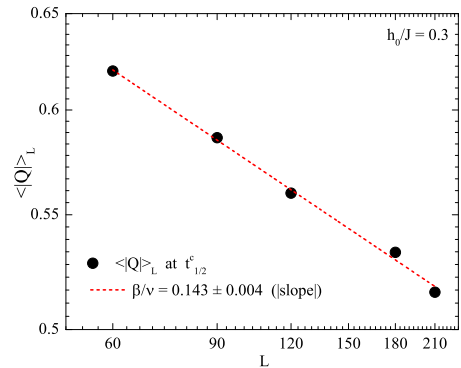}
\caption{(Color online) Log-log plot of the dynamic
order parameter $\langle |Q|\rangle_{L}$ as a function of the linear
system size $L$ for the kinetic Ising model on a triangular lattice
at $t_{1/2}=t_{1/2}^c$. We note that the filled symbols denote the numerical
data obtained from MC simulation while the red line is the
weighted least square fit. The numerical data are collected by
averaging over 200 000 periods of the magnetic field.}\label{Fig6}
\end{figure}

As we noted before, by means of Eq. \ref{Eq6} and \ref{Eq7}, it is
possible to determine the critical exponents of the kinetic Ising model
on a triangular lattice.  We give log-log plot of the dynamic
order parameter $\langle |Q|\rangle_{L}$ as a function of the linear
system size $L$ at $t_{1/2}=t_{1/2}^c$  in Fig. \ref{Fig6}. The obtained
simulation findings estimate that the critical exponent
is $\beta/\nu=0.143\pm0.004$ for the dynamic order parameter.

\begin{figure}[h!]
\center
\includegraphics[width=6cm]{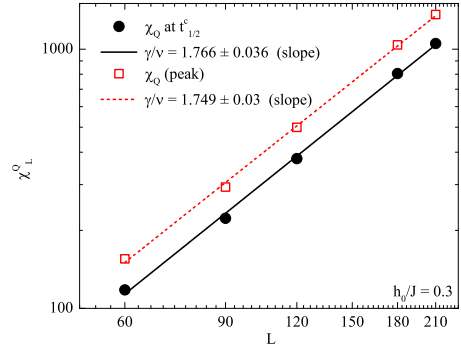}
\caption{(Color online) Log-log plot of the susceptibility $\chi_{L}^{Q}$ as a
function of the linear system size $L$ for the kinetic Ising model on a
triangular lattice. We note that the symbols
denote the numerical data obtained from MC simulation while the lines are the
weighted least square fits. The numerical data are collected by
averaging over 200 000 periods of the magnetic field.}\label{Fig7}
\end{figure}

As a final investigation, we obtain the critical exponent $\gamma/\nu$ by benefiting
from the slopes  of the log-log plot of  the
susceptibility $\chi_{L}^{Q}$ as a function of the system size. It has been found
that the critical exponents are $\gamma/\nu=1.766\pm0.036$
(using the data obtained at $t_{1/2}^c$) and $\gamma/\nu=1.749\pm0.03$ (using
the data at the peak location). It is interesting to note that our
estimates on the critical exponents of the kinetic Ising model
on a 2D triangular lattice are very close to those of
the 2D equilibrium Ising model, which are $\beta/\nu=1/8=0.125$
and $\gamma/\nu=7/4=1.75$.  With the present study,
it is possible to underline that the symmetry arguments reported in Ref. \cite{Grinstein} is also valid
for the 2D triangular lattice under presence of a square-wave magnetic field
considered here, in addition to the previously
published studies \cite{Sides1, Sides2, Korniss1, Buendia, Park1}.

\section{Concluding Remarks}\label{Conclusion}
In this study, we have investigated the magnetic response of the kinetic Ising model on a
2D triangular lattice to a square-wave magnetic field. We have performed
Monte Carlo simulation with single site update Metropolis algorithm. Our numerical
findings clearly indicate that the present system undergoes a DPT at the critical
half-period of the external magnetic field $t_{1/2}^c=142\pm1$ MCSS. It is has been found
that for large half-period of the magnetic field $(t_{1/2}\gg t_{1/2}^c)$ time dependent
magnetization can be capable of following the external field with a relatively small phase lag,
which indicates the dynamically disordered phase. However, for small values of   half-period of the
external fields $(t_{1/2}\ll t_{1/2}^c)$, magnetization does not have enough time to follow
the external magnetic field, and it oscillates around a finite value corresponding to the dynamically
ordered phase.

Moreover, we focus our attention on the finite-size scaling analysis and critical exponents of the
present system, by changing the system size ranging from $L=60$ to $210$. Note that critical
exponents within the statistical errors obtained in this study are found to be consistent with the
universality class of the 2D equilibrium Ising model, as in the case of the previously
published studies \cite{Sides1, Sides2, Korniss1, Buendia}. It seems to be that kinetic spin
models without surfaces may have the same  critical exponents with the corresponding equilibrium Ising model.
However, there exists a few systematic studies done in this direction. Hence, much more work is
required to have better understanding of the DPTs and to classify
universality properties of the spin system far from equilibrium.

\section*{Acknowledgements}
The author is thankful to P.A. Rikvold from Florida State University for useful 
suggestions. The numerical calculations reported in this paper were performed at T\"{U}B\.{I}TAK ULAKBIM (Turkish agency),
High Performance and Grid Computing Center (TRUBA Resources).

\end{document}